\documentclass{ws-rv9x6}  

\begin{document}

\setcounter{chapter}{0}

\chapter{NONEXTENSIVE ENTROPY APPROACH TO\\
SPACE PLASMA FLUCTUATIONS AND TURBULENCE}

\markboth{Leubner V\"{o}r\"{o}s and Baumjohann}{Nonextensive entropy
and space plasma turbulence}

\author{M. P. Leubner\cite{1}, Z. V\"{o}r\"{o}s\cite{2} and W. Baumjohann\cite{2}}

\address{\cite{1}Institute of Astrophysics, University of Innsbruck, Austria\\
E-mail: manfred.leubner@uibk.ac.at\\
\cite{2}Space Research Institute, Austrian Academy of Sciences, Graz, Austria}

\begin{abstract}
Spatial intermittency in fully developed turbulence is an established
feature of astrophysical plasma fluctuations and in particular apparent
in the interplanetary medium by in situ observations. In this situation the
classical Boltzmann-Gibbs extensive thermo-statistics, applicable when
microscopic interactions and memory are short ranged and the environment
is a continuous and differentiable manifold, fails. Upon generalization 
of the entropy function to nonextensivity,
accounting for long-range interactions and thus for correlations in the
system, it is demonstrated that the corresponding probability distributions
(PDFs) are members of a family of specific power-law distributions. In particular,
the resulting theoretical bi-kappa functional reproduces accurately the
observed global leptokurtic, non-Gaussian shape of the increment PDFs
of characteristic solar wind variables on all scales, where nonlocality
in turbulence is controlled via a multi-scale coupling parameter. Gradual
decoupling is obtained by enhancing the spatial separation scale
corresponding to increasing kappa-values in case of slow solar wind
conditions where a Gaussian is approached in the limit of large scales.
Contrary, the scaling properties in the high speed solar wind are
predominantly governed by the mean energy or variance of the
distribution, appearing as second parameter in the theory. 
The PDFs of solar wind scalar field differences are computed
from WIND and ACE data for different time-lags and bulk speeds and analyzed
within the nonextensive theory, where also a particular nonlinear dependence
of the coupling parameter and variance with scale arises for best fitting
theoretical PDFs. Consequently, nonlocality in fluctuations, related to both,
turbulence and its large scale driving, should be related to  
long-range interactions in the context of nonextensive entropy
generalization, providing fundamentally the physical background of the
observed scale dependence of fluctuations in intermittent space plasmas.
\end{abstract}

\section{Introduction}     

Leptokurtic, long-tailed probability distribution functions (PDF's)
subject to a non-Gaussian core and pronounced halo are a persistent
feature in a variety of different astrophysical
environments. Those include the thermo-statistical properties of
the interplanetary medium where the electron, proton and even heavy
ion velocity space distributions show ubiquitously suprathermal halo
patterns (see Mendis\cite{1} for a general review or Leubner\cite{2,3}
and references therein), well described by the empirical family
of kappa-distributions, a power law in particle speed, recognized first by
Vasyliunas\cite{4}. In continuation, significant
progress was provided by Treumann\cite{5,6} who developed
a kinetic theory, demonstrating that power-law velocity space distributions
are a particular thermodynamic equilibrium state.
Similarly, scale invariant power-law distributions
are manifest in any systems relying on self organized criticality (SOC)
(Bak\cite{7}, Bak et al.\cite{8}, Watkins et al.\cite{9}, 
Chapman etal.\cite{10}, Chapman and Watkins\cite{11}) or gravitationally
bound astrophysical stellar systems (Nakamichi et al.\cite{12},
Chavanis and Bouchet\cite{13}).
Moreover, recently Leubner \cite{14} developed a theory representing accurately
the hot plasma and dark matter density profiles in galaxies and clusters
in the context of scale invariant power-law distributions. In all cases the
standard Boltzmann-Gibbs-Shannon (BGS) statistics does not apply . 

Remarkably, we have to add to this diversity also the PDF's of the turbulent
fluctuations of the magnetic field strength, density and velocity fields
in space and astrophysical plasmas (Leubner and V\"{o}r\"{o}s\cite{15,16}). In
particular, the analysis of the PDFs of the solar wind plasma is of
considerable interest to study intermittency and multi-scale statistical
properties in fully developed turbulence, since high resolution in situ
observations are available. The characteristics of the spectral properties
of fluctuations in the incompressible interplanetary medium were provided
in classical statistical theory via the phase space distribution obtained
from ideal MHD invariants by Matthaeus and Goldstein\cite{17} and followed by a
confirmation of the existence of solar wind multifractal structures
({Burlaga\cite{18,19}). Furthermore, solar wind observations
were also able to study the differences between fluid and MHD turbulence 
(Carbone\cite{20}).

The non-Gaussianity of the PDFs of the magnetic field and plasma
fluctuations was analyzed and linked to intermittency and the
fractal scaling of the solar wind MHD fluctuations (Marsch and Tu\cite{21,22}),
followed by detailed investigations of the non-Gaussian fractal and/or
multifractal characteristics of solar wind and related magnetospheric
parameters (Hnat\cite{23}). Their multiscale coupling properties
and significance in view of magnetospheric response was
analized (V\"{o}r\"{o}s  et al.\cite{24}, V\"{o}r\"{o}s and
Jankovi\v{c}ov\'{a}\cite{25}), including studies of multi-scale
intermittency and anisotropy effects in the near-Earth magnetotail
dynamics (V\"{o}r\"{o}s et al.\cite{26,27}). 

WIND, ACE and Voyager observations of solar wind multi-scale
statistical properties verify that the leptokurtic, long-tailed
shapes of the PDFs at small scales represent the characteristics of
intermittent turbulence and approach a Gaussian, reflecting a
decoupled state, on large scales (Sorriso-Valvo et al.\cite{28},
Burlaga et al.\cite{29}, Burlaga et al.\cite{30}).
In other words, the probability of rare events is raised on small scales,
where the spatial separation scale is characterized commonly by the
differences $\delta X(t) = X(t + \tau) - X(t)$, $X(t)$ denoting any
characteristic solar wind variable at time $t$ and $\tau$ is the time lag.
Recently, intermittency was considered to appear as result of an 
interplay between stochastic Alfv\`{e}nic fluctuations and coherent
2-D structures (Bruno et al.\cite{31})

The empirical  Castaing model (Castaing et al.\cite{32}, 
Castaing and Dubrulle\cite{33}, Castaing\cite{34})
introduces intermittency through fluctuations of log-normal 
distributed variances based on the idea that for constant energy
transfer between spatial scales all variables obey a Gaussian
distribution of fluctuations $\delta X$ and hence assuming that
fluctuations on different scales are independent. 
This convolution of Gaussians of different variances was introduced
to model the non-Gaussian energy cascade character of
intermittency in turbulent flows (Consolini and Michelis\cite{35},
Guowei et al.\cite{36}, Sorriso-Valvo\cite{28}, Schmitt and Marsan\cite{37})
where the log-normal distribution of variances through the inertial scales 
was found to provide excellent fits to the observed leptokurtic PDFs
in solar wind flows. Due to the fitting accuracy the Castaing model
achieved high popularity, but appears to be subject to two
significant shortcomings on physical grounds: the model provides 
(1) no link to non-locality and long-range interactions present
in turbulence and (2) no justification for direct energy coupling
between separated scales. Other known shortcomings of Castaing
or log-normal models are related to the predicted features of
high-order moments, which violate the fundamental assumptions
of turbulence theory (Frisch\cite{71}, Arratia et al.\cite{72}).

The global leptokurtic non-Gaussian shape of the increment PDFs requires
theoretically a corresponding unique global distribution function.
This condition can be formulated on a general level by considering the
basic feature of turbulent flows, i.e. multi-scale coupling
or nonlocality in physical or in Fourier space, where nonlocality appears
due to the presence of long-range forces implying direct nonlocal
interactions between large scales and small scales. Due to long-range
interactions small and large scales are strongly coupled indicating that small-scale
fluctuations in each time/space point depend on the large scale motions
in the whole time/space domain and vice versa (Tsinober\cite{38}).
Accounting for long-range interactions is a particular feature of
nonextensive systems and available from pseudo-additive entropy generalization.

The classical Boltzmann-Gibbs extensive thermo-statistics constitutes a
powerful tool when microscopic interactions and memory are short ranged
and the environment is an Euclidean space-time, a continuous and
differentiable manifold. However, in the present situation we
are dealing with astrophysical systems, generally subject to spatial
or temporal long-range interactions evolving in a non-Euclidean, for
instance multi-fractal space-time that makes their behavior nonextensive.
A suitable generalization of the Boltzmann-Gibbs-Shannon entropy for
statistical equilibrium was first proposed by Renyi\cite{39} and 
subsequently by Tsallis\cite{40}, preserving the usual properties
of positivity, equiprobability and irreversibility, but suitably extending
the standard extensivity or additivity to nonextensivity. The main theorems
of the classical Maxwell-Boltzmann statistics admit profound generalizations
within nonextensive statistics (sometimes referred to as q-statistics where
$q$ characterizes the degree of nonextensivity of the system), wherefore
a variety of subsequent analyses were devoted to clarify the mathematical
and physical consequences of pseudo-additivity, for an early review
see e.g. Tsallis\cite{41}. Those include a reformulation of the classical
N-body problem within the extended statistical mechanics (Plastino et al.\cite{42})
and the development of nonextensive distributions (Silva et al.\cite{43},
Almeida\cite{44}) where a deterministic connection between the generalized entropy and
the resulting power-law functionals ({Andrade et al.\cite{45}), as well as the
duality of nonextensive statistics were recognized (Karlin et al.\cite{46}).

Relating the parameters $q$ and $\kappa$ by the transformation
$\kappa = 1/(1-q)$ (Leubner\cite{47}) provided the missing link between nonextensive
distributions and $\kappa$-functions favored in space plasma physics,
leading to the required theoretical justification for the use of
$\kappa$-distributions from fundamental physics. Since the
parameter $\kappa$, a measure of the degree of nonextensivity of the
system, is not restricted to positive values in the nonextensive context,
the commonly observed core-halo twin character of the interplanetary
electron and ion velocity space distributions was verified
theoretically upon generalization to a bi-kappa distribution, subject
to a less pronounced core along with extended tails, as compared to
a Maxwellian (Leubner\cite{48,49}).

Recently, the PDF of the Tsallis ensemble was linked to the analysis
of fully developed turbulence providing a relation between the
nonextensive parameter $q$ and the intermittency exponent $m$ that
is, a manifestation of multifractality of the distribution of eddies
(Arimitsu and Arimitsu\cite{50}, Arimitsu and Arimitsu\cite{51})
as well as of scaling of the velocity structure functions
(Arimitsu and Arimitsu\cite{52}). Moreover, the context of generalized
thermo-statistics provides analytical formulas for PDFs of distance
dependent velocity differences, linking the entropic index to the cascade
like structure of the turbulent dynamics (Beck\cite{53}). We relate in the
following nonlocality in turbulent flows to the presence of long-range
forces in nonextensive systems and demonstrate in the context of entropy
generalization the consistency of the theoretically derived bi-kappa distribution
(Leubner\cite{48}, Leubner and V\"{o}r\"{o}s\cite{15}) with observed,
scale dependent PDFs of characteristic variables in the intermittent,
turbulent solar wind, where both, slow and high speed conditions are
analyzed separately.

\section{Theory}

The standard BGS statistics is based on the extensive entropy measure

\begin{equation}
S_{B}=-k_{B}{\sum p_{i}}\ln p_{i} 
\label{1}
\end{equation}

where $p_{i}$ is the probability of the $i^{th}$ microstate, $k_{B}$ is Boltzmann's
constant and $S_B$ is extremized for equiprobability. As physical background
one assumes that particles move independently from each other, i.e. there are
no correlations present in the system considered. This implies isotropy of
the velocity directions and thus the entropy appears as additive quantity
yielding the standard Maxwellian distribution function. In other words,
microscopic interactions are short ranged and we are dealing with an
Euclidean space time. The assumptions behind standard BGS statistics are
not applicable if one needs to account for nonlocality and long-range
interactions in a fractal/multifractal physical environment. It is
required to introduce correlation within the system, which is done
conveniently in the context of nonextensive entropy generalization
leading to scale-free power-law PDFs.

Considering two sub-systems $A$ and $B$ one can illuminate nonextensivity
by the property of pseudo-additivity of the entropy such that

\begin{equation}
S_{\kappa}(A+B)=S_{\kappa}(A)+S_{\kappa}(B)+\frac{1}{\kappa} S_{\kappa}(A)S_{\kappa}(B)
\label{2}
\end{equation}

where the entropic index $\kappa$ is a parameter quantifying the degree of
nonextensivity in the system. For $\kappa = \infty$ the last term on
the right hand side cancels leaving the additive entropy of standard BGS
statistics. Hence, nonlocality or long-range interactions are introduced
by the multiplicative term accounting for correlations between the
subsystems. As a measure for entropy mixing the entropic index $\kappa$
quantifies the degree of nonextensivity in the system and thus accounts
for nonlocality and long-range interactions or coupling and correlations, 
respectively.
In general, the pseudo-additive, $\kappa$-weighted term
may assume positive or negative definite values indicating a nonextensive
entropy bifurcation. Obviousely, nonextensive systems are subject to a
dual nature since positive $\kappa$-values imply the tendency to
less organized states where the entropy increases whereas negativ   
$\kappa$-values provide a higher organized state of decreased entropy,
see Leubner\cite{14}. 

The general nonextensive entropy consistent with Eq. (\ref{2}), replacing
the classical BGS-statistics for systems subject to long-range interactions,
takes the form (Tsallis\cite{40}, Leubner\cite{48}) 

\begin{equation}
S_{\kappa }=\kappa k_{B}({\sum }p_{i}^{1-1/\kappa }-1)
\label{3}
\end{equation}

In order to link the $\kappa$-notation defined within 
$-\infty \leq \kappa \leq +\infty$, commonly applied in space
plasma modeling in terms of the family of $\kappa$-distributions,
to the Tsallis q-statistics one may perform the transformation
$1/(1-q) = \kappa$ to Eq. (\ref{3}) (Leubner\cite{47}).
$\kappa = \infty$ corresponds to $q = 1$ and represents the extensive
limit of statistical independency. Consequently, the interaction term in
Eq. (\ref{2}) cancels recovering with respect to Eq. (\ref{3}) the
classical Boltzmann-Gibbs-Shannon entropy Eq. (\ref{1}).
Eq. (\ref{3}) applies to systems subject to long-range interactions or
memory and systems evolving in a non-Euclidean and multifractal space-time.
A further generalization of Eq. (\ref{2}) for complex systems, composed
of an arbitrary number of mutually correlated systems, is provided by 
Milovanov and Zelenyi\cite{54}, where appropriate higher order terms in the entropy
appear. Once the entropy is known the corresponding probability distributions
are available.

In Maxwells derivation the velocity components of the distribution $f(v)$
are uncorrelated where $ln f$ can be expressed as a sum of the logarithms
of the one dimensional distribution functions. In nonextensive systems one
needs to introduce correlations between the components accounting
for the long-range interactions, which is conveniently done by extremizing the
entropy under conservation of mass and energy yielding the corresponding
one-dimensional power-law distributions as

\begin{equation}
f^{\pm}=A^{\pm}\left[ 1+\frac{1}{ \kappa }\frac{v^{2}}{v_{t}^{2}}\right]
^{-\kappa }
\label{4}
\end{equation}

where $v_{t}$ corresponds to the mean energy or thermal speed of the distribution.
Hence, the exponential probability function of the Maxwellian gas of an
uncorrelated ensemble of particles is replaced by the characteristics of
a power law where the sign of $\kappa$, indicated by superscripts, governs
the corresponding entropy bifurcation. 
Note that the distribution (\ref{4}) can be derived entirely
using general methods of statistical mechanics without introducing
a specific form for long-range interactions. Incorporating the
sign of $\kappa$ into Eq. (\ref{4}) generates a dual solution with
regard to positive and negative $\kappa$-values, respectively,
resulting also in two different normalizations $A^{\pm}$.

The entropy bifurcation appears to be manifested also in higher order
moments yielding for the second moments $\kappa$-dependent
generalized temperatures, see {Leubner\cite{48}. Furthermore,
the positive solution is restricted to $\kappa > 3/2$ whereas the negative
solutions are subject to a cutoff in the distribution at 
$v_{max}= v_{t}\sqrt(\kappa)$.
Both functions, $f^{+}$ and $f^{-}$ in Eqs. (\ref{4}) approach
the same Maxwellian as $\kappa \rightarrow \infty$. Fig. 1, left panel, demonstrates
schematically the non-thermal behavior of both, the suprathermal halo
component and the reduced core distribution, subject to finite
support in velocity space, where the case $\kappa =\infty $
recovers the Maxwellian equilibrium distribution.

Any unique and physically relevant nonextensive PDF must obey
the following three conditions: (a) the distribution approaches
one and the same Maxwellian as $\kappa \rightarrow \infty$, (b) a unique,
global distribution must be definable by one single density and a unique
temperature and (c) upon variation of the coupling parameter $\kappa$
particle conservation and adiabatic evolution are required,
such that a redistribution in a box (a source free environment)
can be performed. Subject to these constraints the appropriate mathematical
functional, representing observed core-halo structures in nonextensive
astrophysical environments, is available from the elementary combination
$f_{ch} = B_{ch}(f_{h} + f_{c})$, $B_{ch}$ being a proper normalization
constant. In this context the full velocity space bi-kappa 
distribution, compatible with nonextensive entropy generalization and 
obeying the above constraints, reads

\begin{equation}
F_{ch}(v;\kappa) = \frac{N}{\pi^{1/2}v_{t}}G(\kappa)
\left\{\left[1+ \frac{1}{\kappa} \frac{v^{2}}{v_{t}^{2}}\right]^{-\kappa} 
+\left[ 1- \frac{1}{\kappa} \frac{v^{2}}{v_{t}^{2}}\right] ^{\kappa }\right\}
\label{5}
\end{equation}

The last term on the right-hand side denotes an expression subject to a thermal
cutoff at the maximum allowed velocity $v_{max}=\sqrt{\kappa}v_{t}$, which
limits also required integrations. Fig. 1 (right panel) indicates also that
the core of the distribution, corresponding to the last term on the right-hand side
of Eq. (\ref{5}), contributes only for small $\delta B$. For details regarding the corresponding
second moments or generalized temperatures see Leubner\cite{48} or
Leubner\cite{49}. The function $G(\kappa)$ is defined by

\begin{equation}
G(\kappa)=\left[ \frac{\kappa^{1/2}\Gamma (\kappa-1/2)}{\Gamma (\kappa)}+
\frac{\kappa^{1/2}\Gamma(\kappa+1)}{\Gamma (\kappa +3/2)}\right]^{-1}
\label{6}
\end{equation}

from the normalization and is subject to a particular
weak $\kappa$-dependence where $G(\kappa) \sim 1/2$, see Leubner\cite{48}
for a graphical illustration and discussion. Hence, the normalization
is independent of the parameter $\kappa$ and the
factor $1/2$ reflects consistently the superposition of the two 
counter-organizing contributions in Eq. (\ref{5}).
For $\kappa = \infty$, $G(\kappa) = 1/2$ and
the power laws in the brackets of the right hand side of Eq. (\ref{5}) turn
each into the same Maxwellian exponential.

\begin{figure}[t]
\vspace*{2mm}
  \centering{\includegraphics[width=5.6cm]{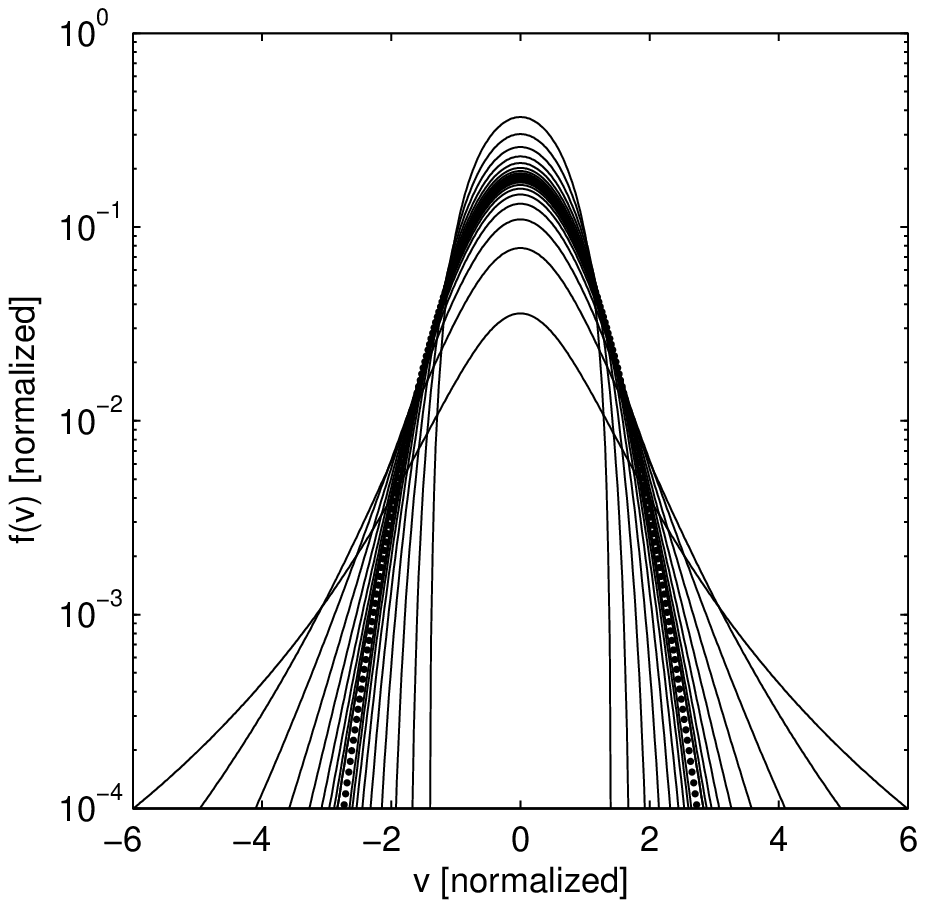}
            \includegraphics[width=5.6cm]{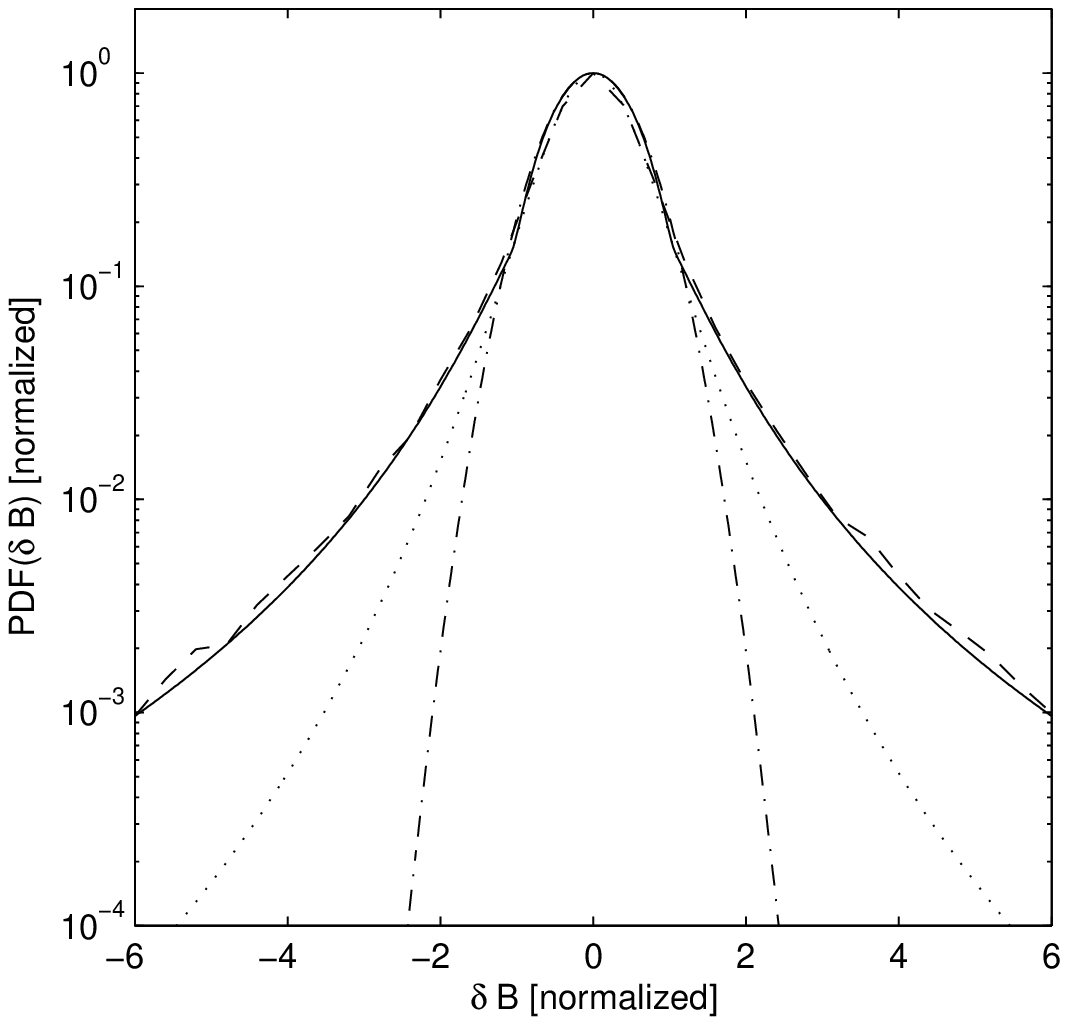}}
  \caption{\label{fig1}
Left panel: A schematic plot of the characteristics of the nonextensive bi-kappa
distribution family: with $\kappa = 3$ the outermost and
innermost curve correspond to the halo $f_{h}$ and core $f_{c}$ distribution
fraction. For increasing $\kappa$-values both sets of curves merge at
the same Maxwellian limit, indicated as bold line, $f_{h}$ from outside
and $f_{c}$ from inside.
Right panel: A nonextensive bi-kappa fit (solid line) with $\kappa=1.8$
of an observed PDF (dashed line) obtained from ACE magnetic field amplitude
data. A Gaussian (dashed-dotted line) and a conventional $\kappa-$distribution
are provided for comparison. }
\end{figure}

With regard to this specific mathematical feature the
energy levels $E_{i}$ of the turbulent spectrum can be related to the
corresponding kinetic energy of velocity differences
$\delta v(t) = v(t + \tau) - v(t)$ between two points of separation $\tau$ 
allowing to transform the 1D Maxwellian particle distributin of mean
energy $v_{t}$ into the mathematical form of a Gaussian of variance
$\sigma$. Upon normalizing the one-dimensional bi-kappa particle
distribution (\ref{5}) to unity and assigning the distribution variance
$\sigma$ to the thermal spread $v_{t}$ the 'Maxwellian form' of the bi-kappa
distribution transforms to a 'Gaussian form' of a global bi-kappa
PDF as

\begin{equation}
P_{ch}(\delta X;\kappa,\sigma)=\frac{1}{2 \sqrt{\pi} \sigma}
\left\{\left[1+ \frac{\delta X^{2}}{\kappa \sigma^{2}}\right]^{-\kappa}  
+\left[ 1- \frac{\delta X^{2}}{\kappa \sigma^{2}}\right] ^{\kappa }\right\}
\label{7}
\end{equation}

This two-parameter PDF is applicable
to the differences of the fluctuations $\delta X(t) = X(t + \tau) - X(t)$ of any
physical variable $X$ in the astrophysical system considered. Here $\kappa$
assumes a physical interpretation defining the degree of nonextensivity
or nonlocality in the system, thus being a measure of the degree of
organization or intermittency, respectively (Leubner and V\"{o}r\"{o}s\cite{15})
and $\sigma$ denotes the distribution variance.
Again, depending on the particular choice of the parameters, the last term
on the right-hand side is subject to a cutoff, treated by limiting the integration
to the proper interval. Fig. 1, left panel, illuminates also that large
values of $\delta X$, corresponding to the tails of the distribution, are
represented by the first term on the right-hand side of Equation (\ref{7}). 
As $\kappa \rightarrow \infty$ the bi-kappa distribution $P_{ch}(\delta X; \kappa, \sigma)$
approaches a single Gaussian. 

Note, that Equation 5 represents the full velocity space bi-kappa distribution,
while Equation 7 is related to the time shifted differences of any physical variale $X$.
Physically, Equation 5 describes particles which motion is controlled by long-range
forces and interactions. The basic assumption for deriving the velocity space
bi-kappa distribution was the pseudo-additivity of the entropy of particle sub-systems
expressed through Equations 2 and 3. It is important to recognize that, the same
type of expression for bi-kappa distribution is obtained, if instead of interacting
particles we assume interacting coherent structures with the same pseudo-additivity
property of the entropy as in Equation 2. In the context of MHD, nonpropagating
multi-scale coherent structures or flux tubes can interact, deform and produce
new sites of nonpropagating fluctuations. Coherent structures of the same polarity
merge into a structure with lower local energetic state, while structures of 
opposite polarities may repel each other (Chang\cite{64}; Bruno and Carbone\cite{69}).
These coherent structures can be considered as discrete interacting "particles"
in MHD flows, responsible for the particular entropy within the system
and validating the analogy to the kinetic level of PDFs (Leubner and V\"{o}r\"{o}s\cite{16}).
A usual way of statistical analysis of intermittency due to the occurrence of coherent
structures in turbulence is through two-point differences of fluctuations.
Therefore, in the expression of bi-kappa distribution for turbulent interactions (Equation 7)
differences of physical variables can be used instead of the velocity notation.  
Moreover, passive scalars as the magnetic field follow the dynamics of
V or $\delta V$ (see V\"{o}r\"{o}s et al.\cite{70}).

One of the basic future of turbulent flows is multi-scale redistribution of energy
during which interacting coherent structures appear, reducing the entropy of the system
and leading also to negative kappa values. At the same time, turbulence enhances dissipation and mixing
of the plasma, which increases entropy and can be described in terms of positive kappa values.
The presence of processes which increase entropy and those which decrease entropy in turbulence
advocates therefore the introduction of bi-kappa distributions.
 
We proof in the following the relevance of the
nonextensive, global bi-kappa PDF (\ref{7}) on the observed scale dependence
of the PDFs of the differences of magnetic field, velocity and density
variables in the intermittent, turbulent interplanetary medium.

\section{Application: Scale dependent interplanetary PDFs}

Based on the nonextensive two parameter bi-kappa
distribution (\ref{7}) we compare the PDFs obtained from slow
and fast solar wind data with particular attention to the scale dependent
changes of the two physically interpretable parameters ($\kappa$, $\sigma$) involved.
We do not consider any occurrence of discontinuities or shocks in the system.
The problem of interaction of turbulence with large scale structures
as shocks is investigated elsewhere. (V\"{o}r\"{o}s et al.\cite{70})

\begin{figure*}[t]
\vspace*{2mm}
  \centering{\includegraphics[width=11cm]{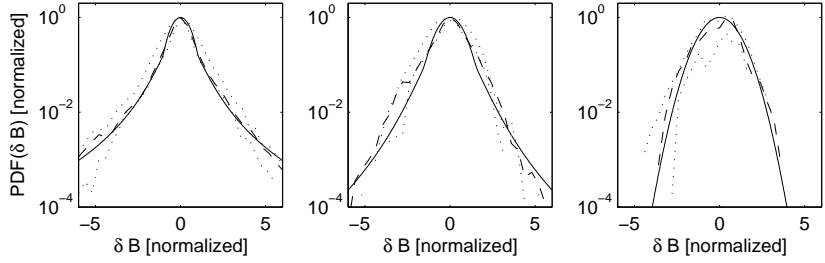}}
  \caption{\label{fig2}
The PDF of the increments of observed ACE magnetic field fluctuations for
$\tau = 100$ and a resolution of 16s as compared to the bi-kappa function with
$\kappa = 1.8$. Based on the same data the central panel provides the
characteristics for increased $\tau = 2000$ where $\kappa$ assumes
a value of 3.0 for the best representation. The PDF of large-scale magnetic
field fluctuations, $\tau = 10000$, are well modeled by a Gaussian with
$\kappa = \infty$, right panel. The dotted lines correspond
to the standard deviations.}
\end{figure*}

For each data set the magnetic field and plasma parameter increments
were calculated at a given time lag $\tau$ by 
$\delta X(t) = X(t + \tau) - X(t)$, where $X$ represents any 
physical variable considered. For each 
realization the empirical probability distribution
function (histogram) was then computed. $\delta X(t)$ is binned into $n$
equal spaced boxes and the number of elements in each box was computed
where the robustness of the histograms against $n$ is tested. $\delta X(t)$
represents characteristic fluctuations at the time scale $\tau$ or, equivalently,
across eddies of size $l \sim v\tau$. Hence, by changing $\tau$ it is possible
to analyze the statistical features of fluctuations in different time scales,
which roughly correspond to those statistical characteristics across
turbulent eddies of size $l \sim v\tau$. 

In the following we demonstrate by means of Eq. (\ref{7})
from first principle statistics that the strong non-Gaussianity of the PDF
of small scale fluctuations should be associated physically with long-range
interactions provided in nonextensive systems by pseudo-additive entropy
generalization. The scale dependence of the PDF in the solar wind can
be represented accurately via the tuning parameters $\kappa$ and $\sigma$ of the
bi-kappa functional. 

In Fig. 1, right panel, we focus on the bi-kappa model demonstrating 
that the nonextensive context, generates a precise representation
for the observed solar wind PDF characterizing the intermittency of the small
scale fluctuations. For comparison also a Gaussian and the conventional
$\kappa-$function (Leubner\cite{47}), subject to the same $\kappa-$value
but not able to reproduce the structure of the PDF of small scale fluctuations,
are provided. 
It is also possible to assume that the smallest fluctuations are random and uncorrelated,
and this might be the reason why the central part of the distribution near the maximum is well-fitted by
a Gaussian, as in the right panel in Fig. 1. Note, however, that the characteristic scale of the 
differences of fluctuations is introduced through the time lag $\tau$. Therefore, 
$\delta B \to 0$ near the maximum of the distribution simply means that, at the scale $\tau$
the differences between the corresponding values of $B(t+\tau)$ and B(t) are very small.
Combined with the finite precision of PDFs estimation, small two-point differences  
can really produce uncorrelated fluctuations near PDFs maxima. This does not mean, 
however, that the Gaussian distribution is the correct answer
for the proper description of the central part of the distribution at the small scales $\tau$.
If it was true, we should suppose that,
at smaller and smaller scales $\tau$, the fluctuations become more and more random and
we are getting closer and closer to a Gaussian distribution. Actually the opposite is true,
the peakedness of the distribution increases as the scale $\tau$ decreases (see later).
\begin{figure*}[t]
\vspace*{2mm}
  \centering{\includegraphics[width=11cm]{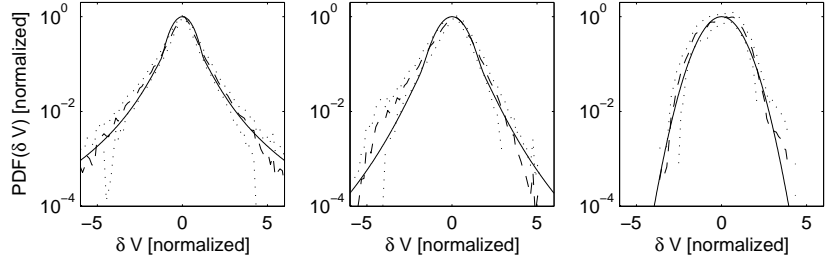}}
  \caption{\label{fig3}
The PDF of the increments of observed WIND velocity field magnitude fluctuations
 for $\tau = 10$ and a resolution of 92s as compared to the bi-kappa function with
$\kappa = 2$. Based on the same data the central panel provides the characteristics
for increased $\tau = 70$ where $\kappa$ assumes a value of $3.5$ for the best
representation. The PDF of large-scale velocity fluctuations are well modeled
by a Gaussian with $\kappa = \infty$, right panel. The dotted lines correspond
to the standard deviations.}
\end{figure*}

\subsection{Slow speed solar wind}

As an example, in Fig. 2
undisturbed solar wind ACE magnetic field amplitude data of $16s$
time resolution are analyzed where the dimensionless $\tau$ is multiplied by
the resolution to generate an effective time-lag. In particular, the scale
dependent PDF evolution of magnetic field fluctuations is subject to a two
point separation scale of $\tau = 100, 2000$ and $10000$. The corresponding best fits
of the bi-kappa distribution are obtained for $\kappa = 1.8, 3$ and $\infty$,
measuring the degree of nonextensivity, or coupling, respectively, through
long-range interactions and the dotted lines refer to the standard deviation.
The accuracy of the bi-kappa
distribution fit demonstrates that non-locality in turbulence,
when introduced theoretically by long-range interactions through the
nonextensive context, generates a precise representation for the observed
PDFs characterizing the intermittency of the fluctuations at all scales.

\begin{figure*}[t]
\vspace*{2mm}
  \centering{\includegraphics[width=11cm]{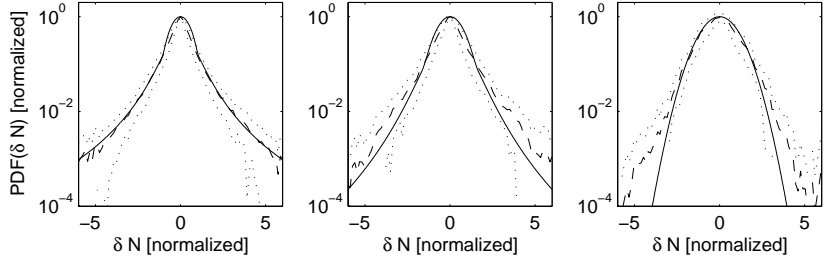}}
  \caption{\label{fig4}
Left panel: The PDF of the increments of observed WIND density
fluctuations for $\tau = 10$ and a resolution of 92 sec.
as compared to the bi-kappa function with $\kappa=2$. Based on the same
data the central panel provides the characteristics for increase $\tau = 70$
where $\kappa$ assumes a value of 3.5 for the best representation. The PDF
of large scale density fluctuations with $\tau=900$ are well modeled by a
Gaussian with $\kappa = \infty$, right panel.}
\end{figure*}

Based on WIND velocity field magnitude data Fig. 3 presents an analysis
of the scale dependence of interplanetary PDFs of the velocity field
magnitude showing in three plots from left to right the decreasing
kurtosis with increasing time-lags,
where the observational uncertainty is again indicated by the standard deviation
(dotted lines). Constraint by a time resolution of $92sec$ the two
point time separation $\tau$ assumes the values $10, 70$ and $900$
from left to right. The solid lines represent best fits to the observed PDFs
with nonextensive distributions, where the corresponding $\kappa$ is determined
as $\kappa = 2, 3.5$ and $\infty$. The strong non-Gaussian character of the
leptokurtic PDFs (left panel), exhibiting pronounced tails associated with
solar wind turbulence and intermittency in small-scale fluctuations, finds
again an accurate analytical fit and hence a physical background in the
nonextensive representation. The non-Gaussian structure is somewhat softened
for enhanced $\tau = 70$ (central panel) but again precisely modeled within
the pseudo-additive entropy context, turning into the Gaussian shape of
large scale fluctuations, which is independent of the increment field.

Fig. 4 provides the corresponding nonextensive analysis of the
scale dependence of the density fluctuations obtained from WIND data
(92 sec resolution). Evidently, the scale dependent characteristics of the
observed PDFs of the increment fields $\delta X(\tau)=X(t+\tau)-X(t)$
for all solar wind variables evolve simultaneously on small scales
approaching independency of the increment field in the large scale
Gaussian. With separation scales of $\tau = 10, 70, 900$ the
corresponding evolution of the PDFs of observed density 
fluctuations are best represented by the same values
of $\kappa = 2, 3.5, \infty$, as for the velocity field
magnitude, accounting for nonlocal interactions
in the nonextensive theoretical approach. 

\begin{figure}[t]
\vspace*{2mm}
  \centering{
  \includegraphics[width=5.6cm]{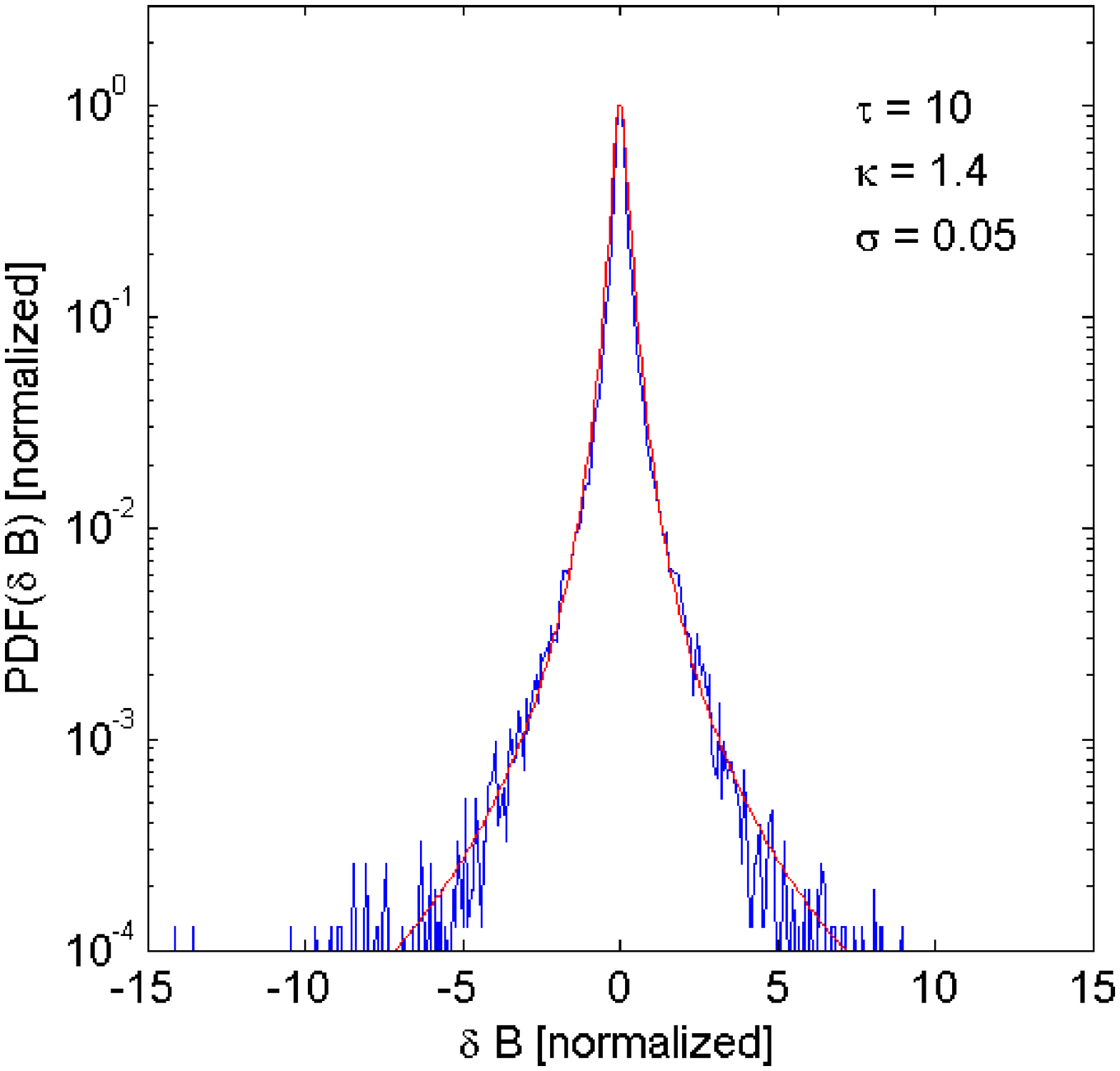}
  \includegraphics[width=5.6cm]{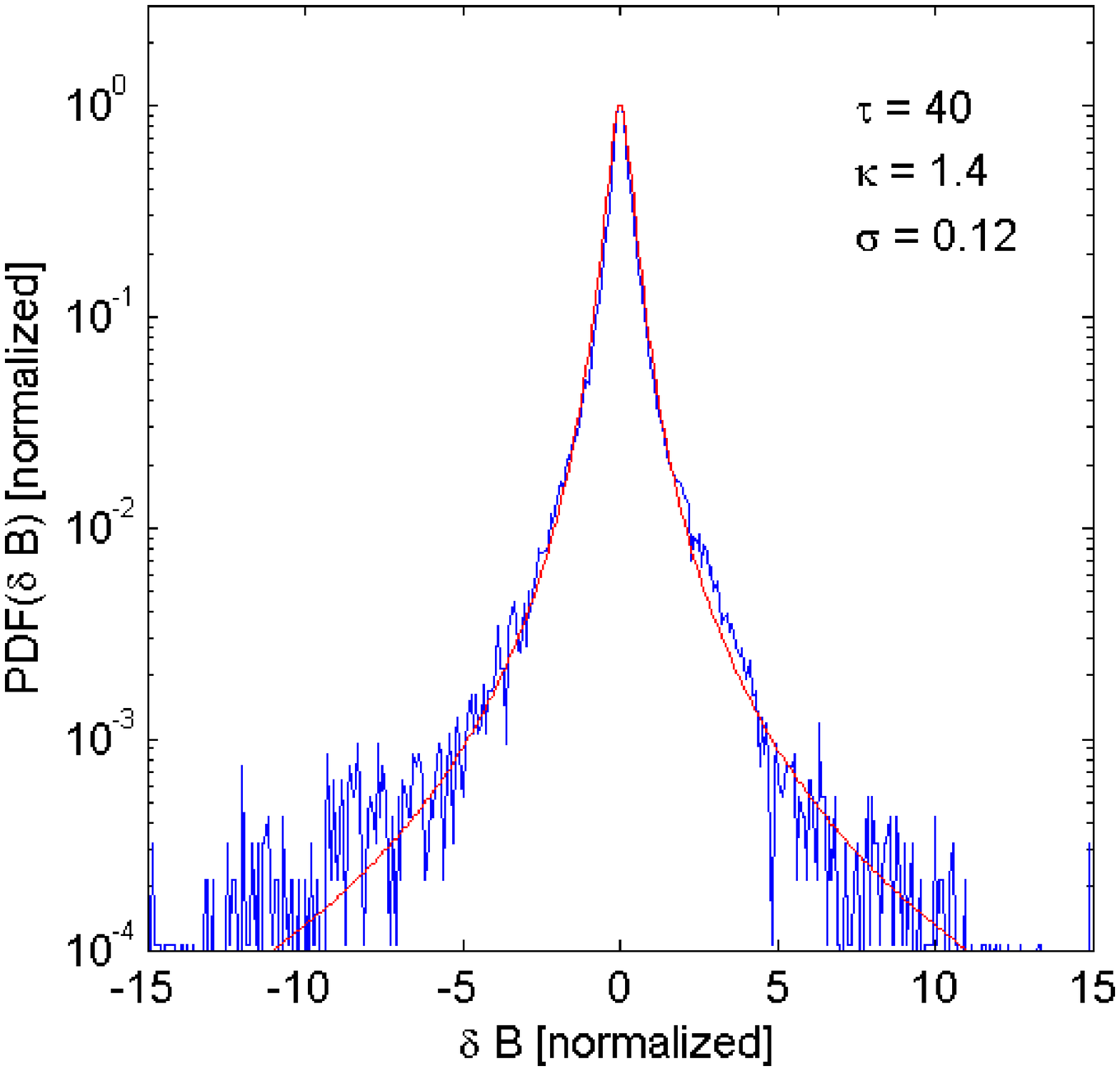}
  \includegraphics[width=5.6cm]{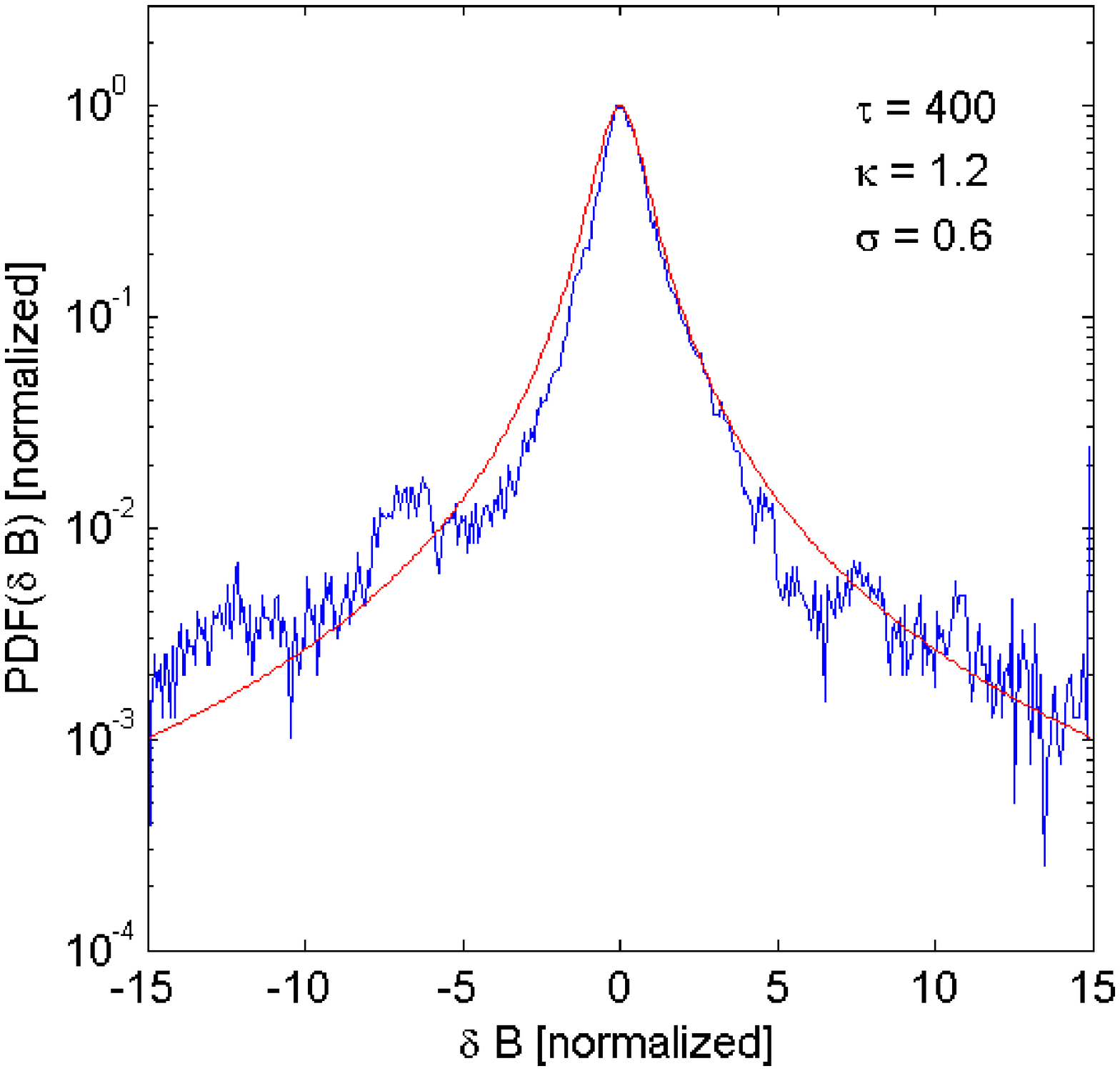}
  \includegraphics[width=5.6cm]{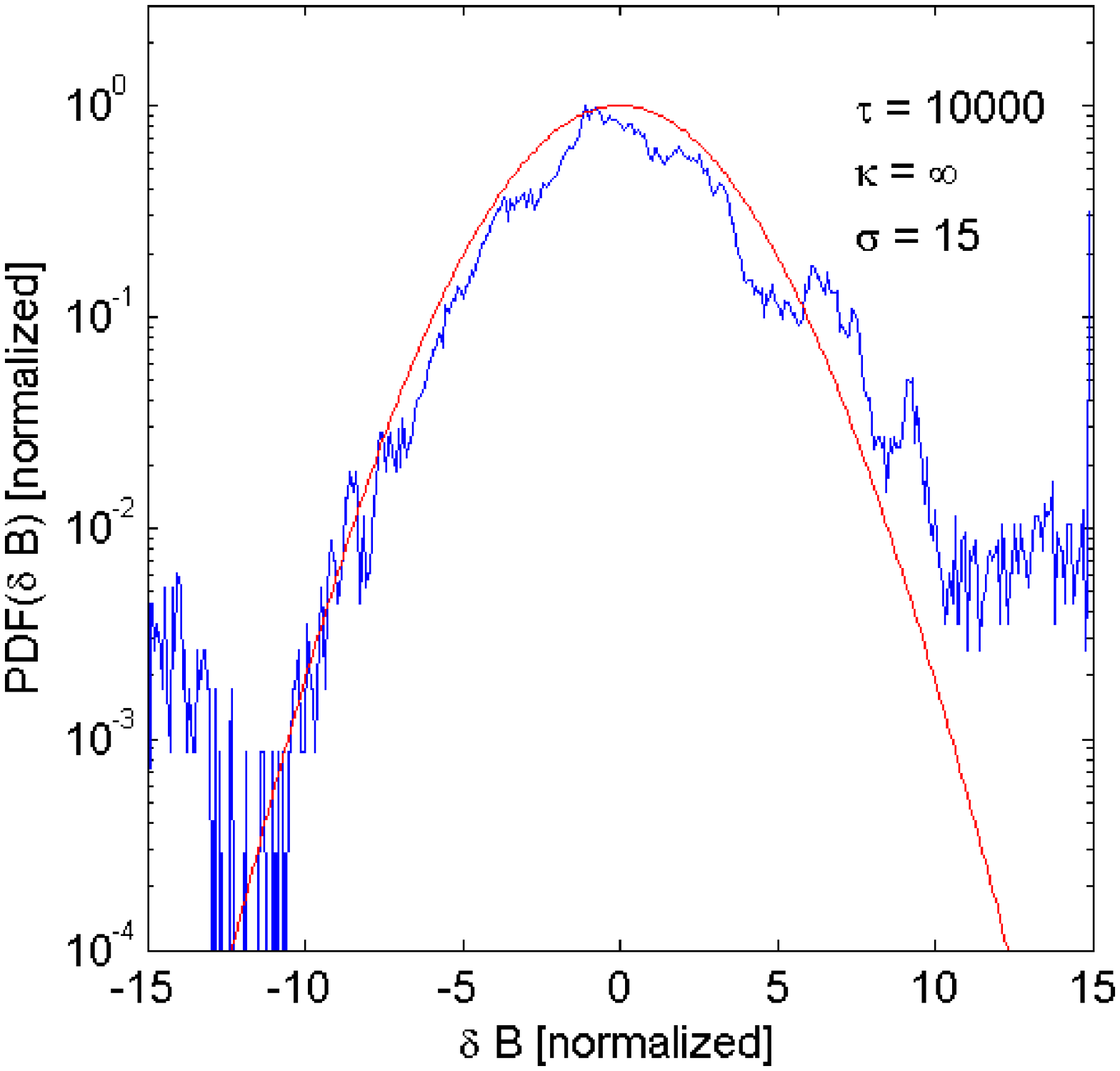}}
  \caption{\label{fig5}
The PDF of the increments of observed ACE high-speed associated magnetic
field magnitude fluctuations (16 s time resolution).
Top-left:  fluctuations at the scale $\tau = 10$ as
compared to the bi-kappa function with $\kappa = 1.4$ and $\sigma=0.05$;
Top-right:  $\tau = 40$, $\kappa = 1.4$ and $\sigma=0.12$; 
Left-bottom: $\tau = 400$, $\kappa = 1.2$ and $\sigma=0.6$;
Left-right: $\tau = 10000$, $\kappa = \infty$ and $\sigma=15$ (Gaussian fit).}
\end{figure}

\begin{figure}[t]
\vspace*{2mm}
  \centering{
  \includegraphics[width=5.6cm]{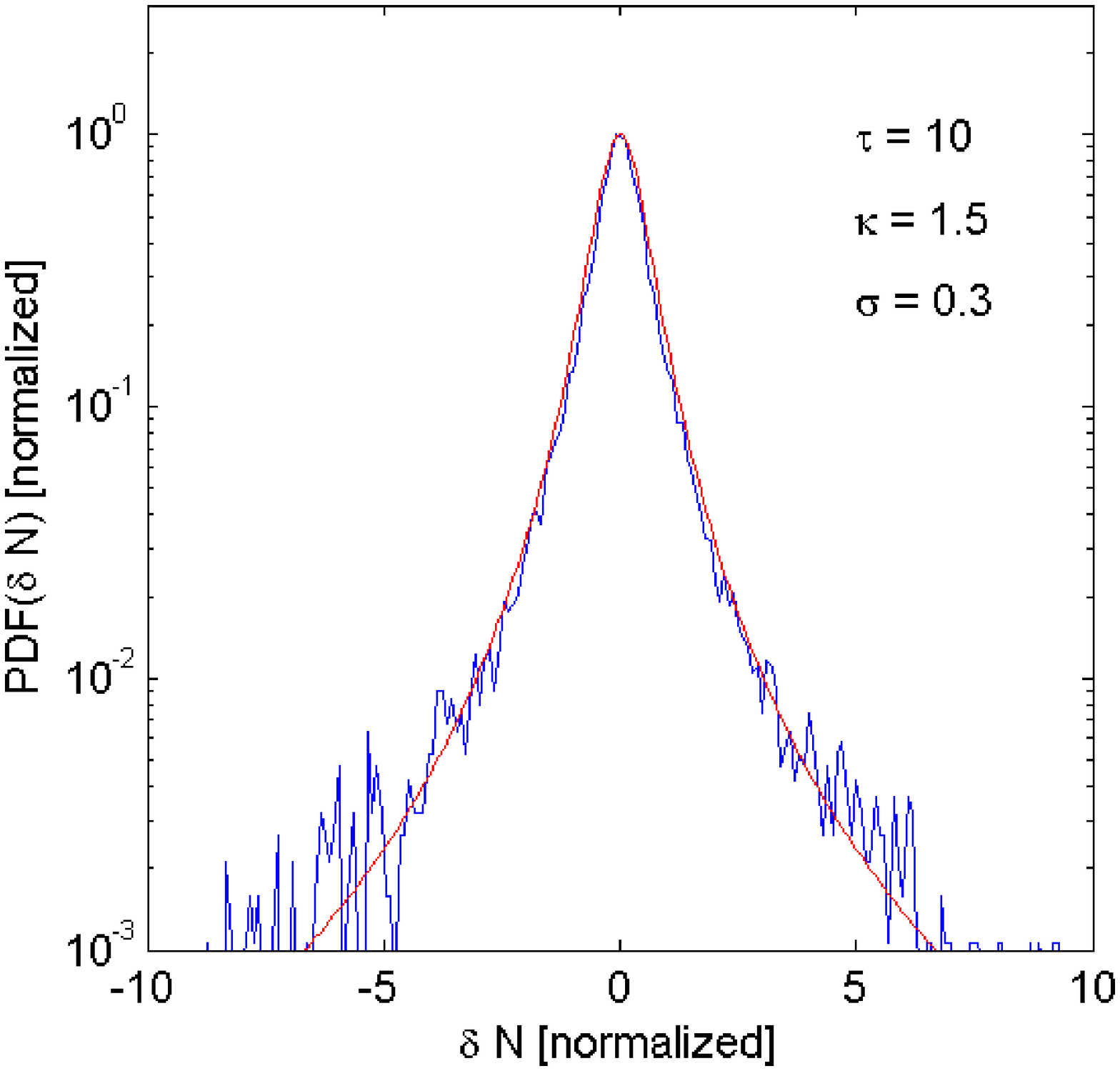}
  \includegraphics[width=5.6cm]{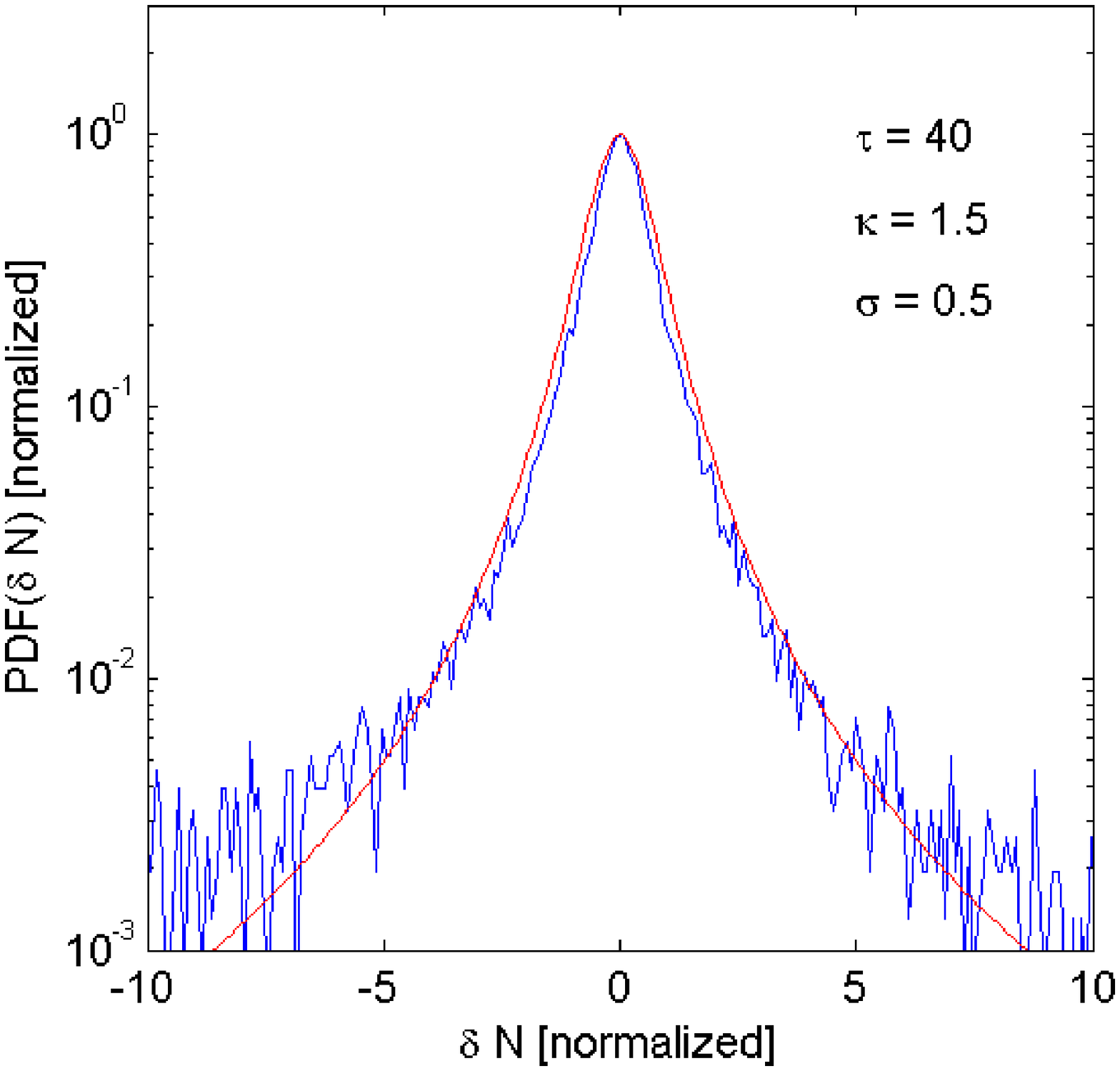}
  \includegraphics[width=5.6cm]{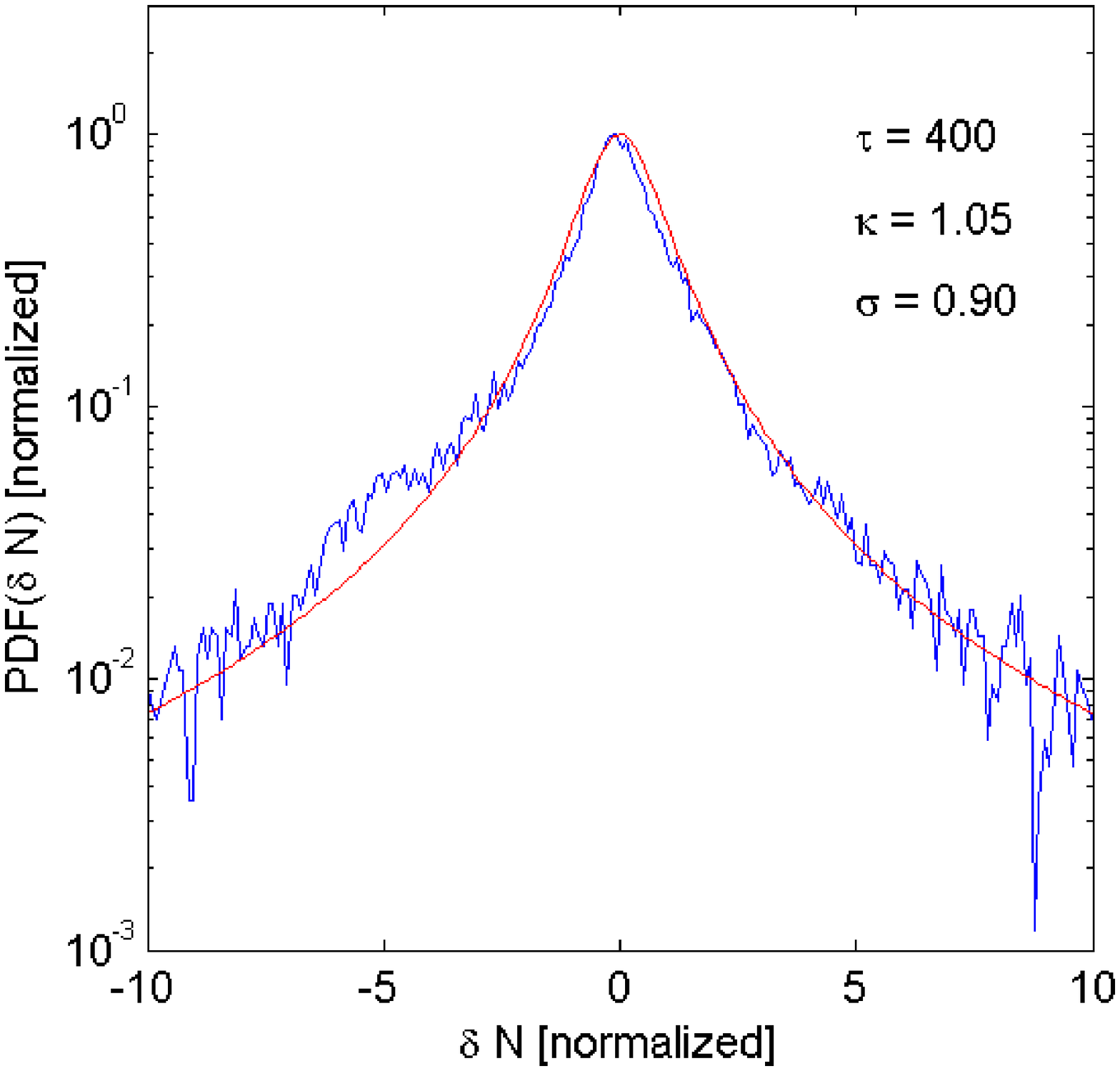}
  \includegraphics[width=5.6cm]{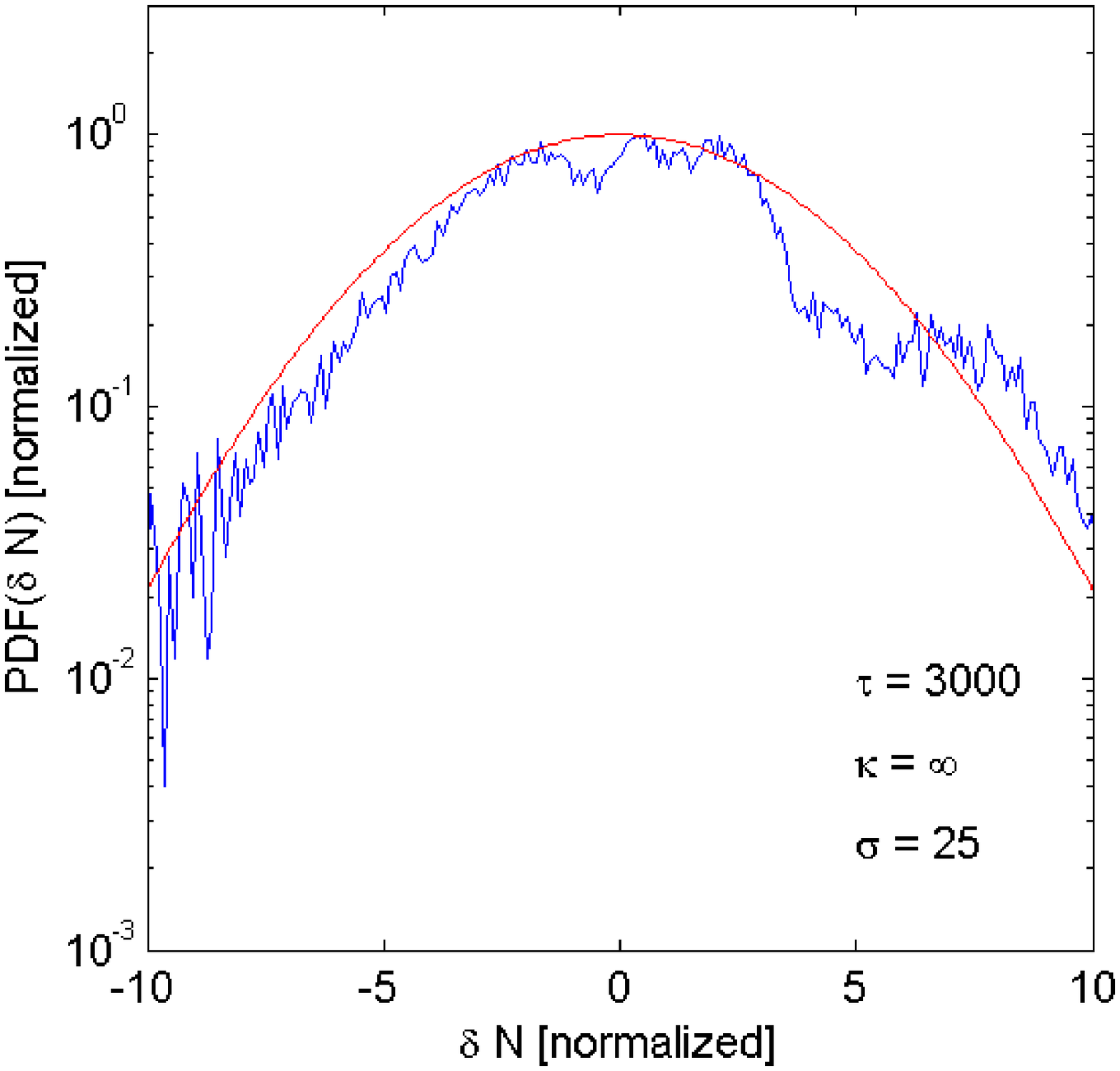}}
  \caption{\label{fig6}
The PDF of the increments of observed ACE high-speed associated density
fluctuations (64 s time resolution).
Top-left:  fluctuations at the scale $\tau = 10$ as compared to the bi-kappa
function with $\kappa = 1.5$ and $\sigma=0.3$; 
Top-right:  $\tau = 40$, $\kappa = 1.5$ and $\sigma=0.5$; 
Left-bottom: $\tau = 400$, $\kappa = 1.05$ and $\sigma=0.9$; 
Left-right: $\tau = 3000$, $\kappa = \infty$ and $\sigma=25$ (Gaussian fit).}
\end{figure}

\subsection{High speed solar wind}

The four panels in Fig.5 show PDFs of high speed associated magnetic field 
magnitude fluctuations. The two-point statistics is demonstrated in the subplots
from top-left to bottom-right for the scales $\tau= 10, 40, 400, 10000$. The
effective time-lag is obtained after multiplying $\tau$ with time resolution
of $16sec$. The  corresponding best fits reveal differing statistical features
of high speed associated magnetic fluctuations. In comparison with
low speed data the degree of nonextensivity does not change during high
speed intervals, $\kappa=1.4, 1.4, 1.2$, over the range of scales
$\tau= 10, 40, 400$, and only for $\tau= 10000$,  $\kappa$ reaches $\infty$.
In contrast to the slow wind data, however, good quality high speed fits  can be
achieved only when the standard deviation $\sigma$ is changed.

It indicates that the abundance of large scale energy content of high speed
flows may facilitate to maintain the degree of nonextensivity and
self-organisation unchanged over the considered scales.
On the other hand, we have no clear explanation yet for the observed changes
of the standard deviation. We can speculate that changes in sigma appear
because of the ample changes in the amplitudes of two-point fluctuations in the
solar wind, having solar origin or being generated  by local processes absent
in the slow wind. Obviously, further comparative case studies on
multi-scale fluctuations are needed to answer the question of the relative
contribution of local processes versus processes originating in the solar corona
to the observed behaviour of statistical moments in the fast and slow solar
wind.

The four panels in Figs. 6 and 7 provide the same qualitative behavior for
high speed associated ACE density ($64sec$ resolution) and WIND magnetic field
($3sec$ resolution).

\begin{figure}[t]
\vspace*{2mm}
  \centering{
  \includegraphics[width=5.6cm]{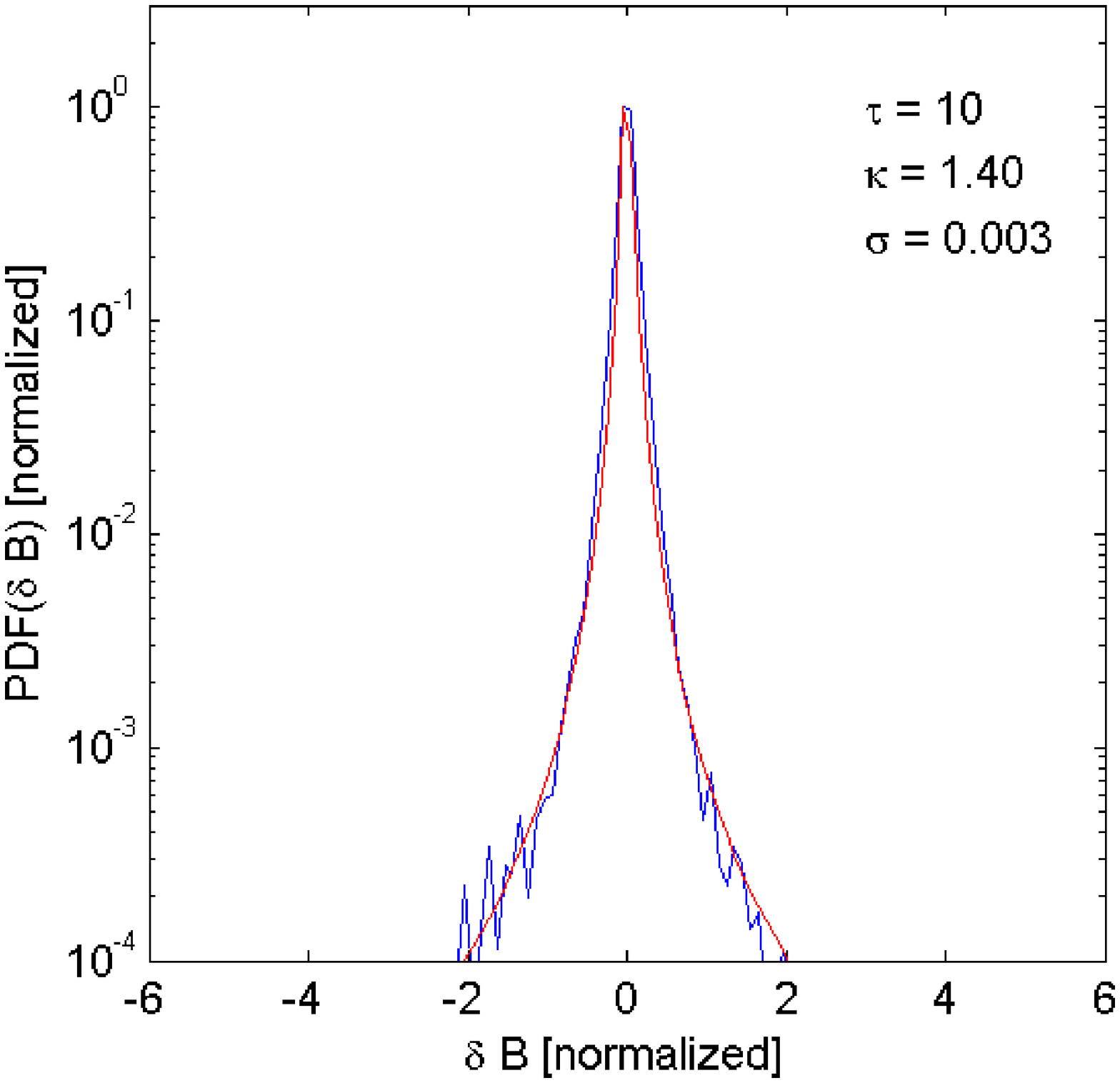}
  \includegraphics[width=5.6cm]{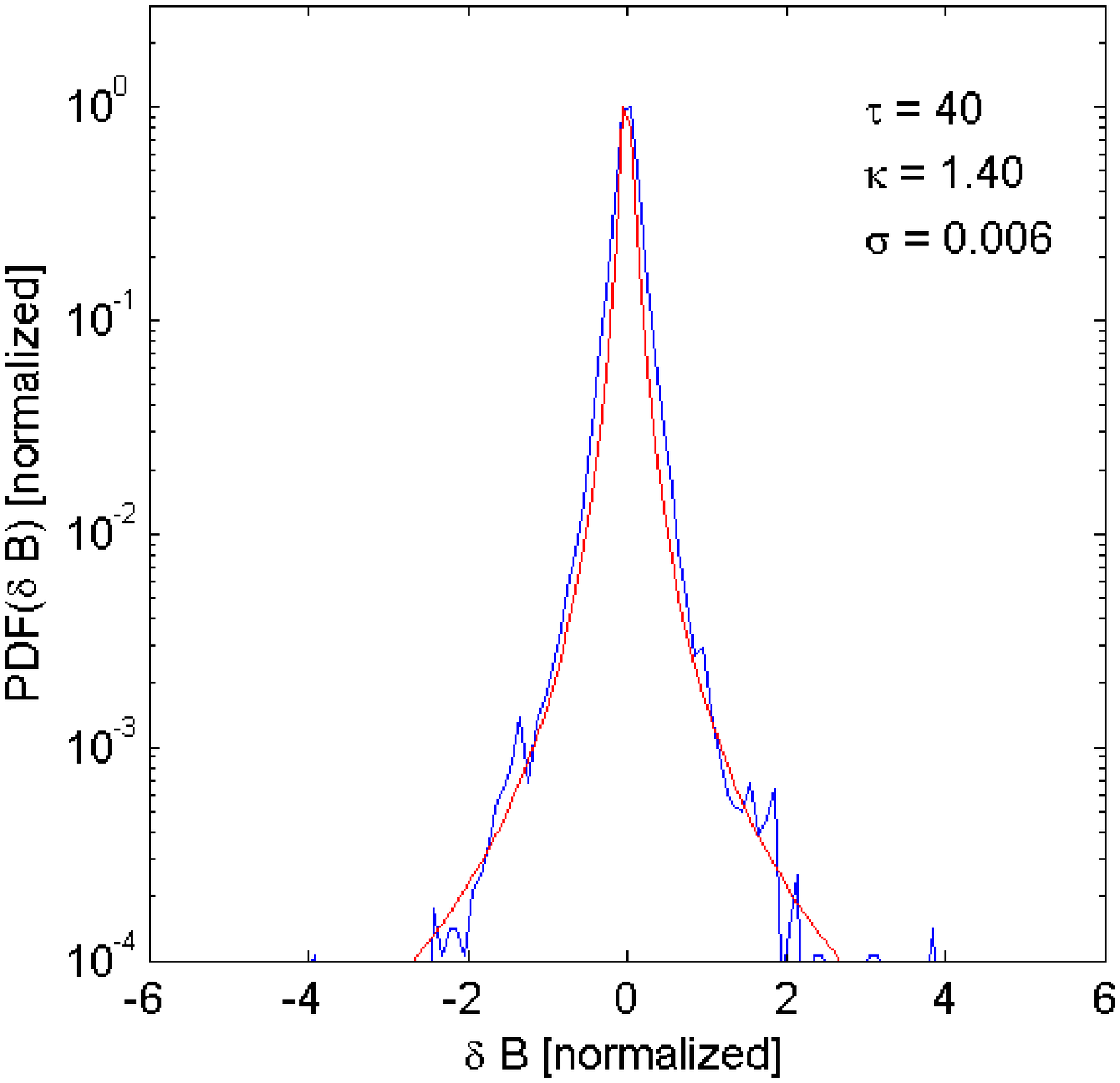}
  \includegraphics[width=5.6cm]{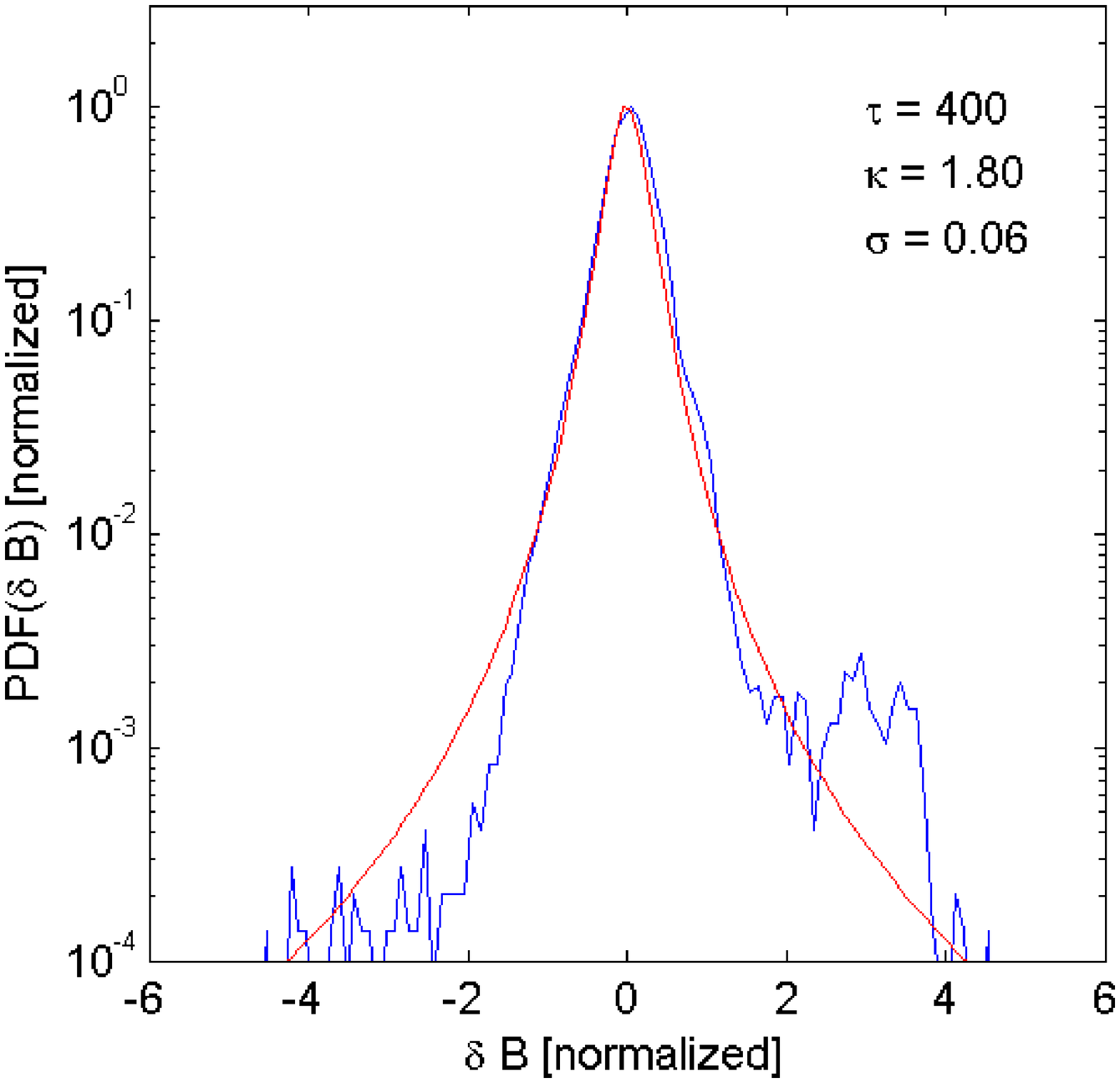}
  \includegraphics[width=5.6cm]{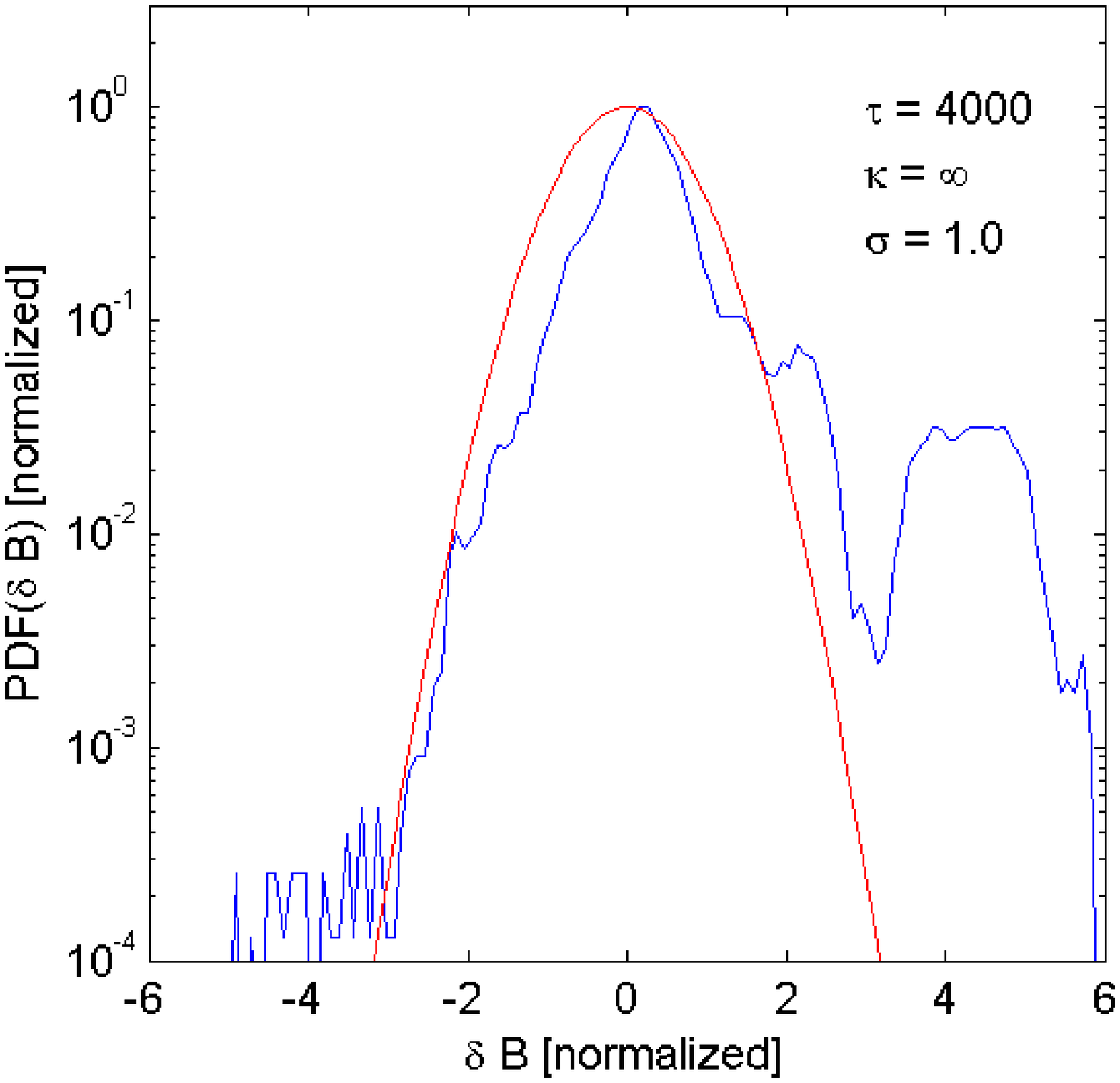}}
  \caption{\label{fig7}
The PDF of the increments of observed ACE high-speed associated magnetic field
magnitude fluctuations (3 s time resolution).
Top-left:  fluctuations at the scale $\tau = 10$ as compared to the bi-kappa
function with $\kappa = 1.4$ and $\sigma=0.003$; 
Top-right:  $\tau = 40$, $\kappa = 1.4$ and $\sigma=0.006$; 
Left-bottom: $\tau = 400$, $\kappa = 1.8$ and $\sigma=0.06$; 
Left-right: $\tau = 4000$, $\kappa = \infty$ and $\sigma=1$ (Gaussian fit).}
\end{figure}

\section{Discussion and conclusions}

The Wind and ACE solar wind data analysis unambiguously manifests that the
PDFs of large scale density, velocity and magnetic field fluctuations are
well represented by a Gaussian, turning into leptokurtic peaked distributions
of strong non-Gaussianity in the center along with a pronounced tail structure
at smaller scales. In particular, the PDFs of large-scale magnetic
field fluctuations, not related to the increment field are known to be
subject to relatively small deviations from the Gaussian statistics and are
well fitted by the Castaing distribution, a convolution of Gaussians with
variances distributed according to a log-normal distribution (Castaing\cite{32},
Padhye et al.\cite{55}). Assuming a constant energy transfer rate between
spatial scales all quantities exhibit a Gaussian distribution of fluctuations
in this context. Independent of the physical situation considered, the
Castaing distribution provides a multi-parameter description of observed PDFs,
plausible in this case, since the large-scale fluctuations of the
interplanetary magnetic field are generated by a variety of discrete
coronal sources. If individual coronal sources evoke Gaussian distributed
magnetic fields, the net magnetic fluctuations can be modeled by their
superposition with a spread of the corresponding variances.

Contrary, small-scale fluctuations are associated with local intermittent
flows where fluctuations are concentrated in limited space volumes.
Consequently, the PDFs are scale dependent and intermittency generates
long-tailed distributions. It is customary to use $n-th$ order absolute powers
of the plasma variables and magnetic field increments ($n-th$ order
structure functions (Marsch and Tu\cite{22}, Pagel and Balogh\cite{56})
allowing to investigate
the multi-scale scaling features of fluctuations.
Direct studies of observed PDFs of the increment fields
$\delta X(t) = X(t + \tau) - X(t)$  for any characteristic solar
 wind variable at time $t$ and time lag $\tau$ revealed departures from a
Gaussian distribution over multiple scales (Sorriso-Valvo\cite{28}) and an increase
of kurtosis (intermittency) towards small scales (Marsch and Tu\cite{21}). The PDFs
are also found to be leptokurtic, which indicates the turbulent character of the
underlying fluctuations. Sorriso-Valvo et al.\cite{28} have shown that the
non-Gaussian behavior of small-scale velocity and magnetic field fluctuations
in the solar wind can also be described well by a Castaing distribution
where the individual sources of Gaussian fluctuations appear at small-scales
in turbulent cascades.

From the corresponding nonextensive WIND data analysis of the density and magnetic
field fluctuations (Leubner and V\"{o}r\"{o}s\cite{15}) it is evident that the scale
dependent characteristics of the observed PDFs of the increment fields
$\delta X(t) = X(t + \tau) - X(t)$  for solar wind variables evolve simultaneously
on small scales, approaching independency of the increment field in the
large- scale Gaussian. Highly accurately, the overall scale dependence
appears as a general characteristic of quiet astrophysical plasma 
environments indicating a universal scaling dependence between density, velocity
and magnetic field intermittency within the experimental uncertainties.  This strong
correlation implies that the scale dependencies of all physical variables are coupled,
where the solar wind Alfv\'{e}nic fluctuations provide a physical basis of the velocity
and magnetic field correlations. On the other hand, according to recent analyses,
the magnetic field intensity exhibits a higher degree of intermittency than the solar
wind bulk velocity, both in fast and slow winds (Sorriso-Valvo et al.\cite{57}). However,
Veltri and Mangeney\cite{58} found that the most intermittent structures in the slow wind
are shock waves, displaying similar intermittency in the magnetic field intensity
and bulk velocity. Furthermore, the proportionality between density fluctuations
and the magnetic field and velocity fluctuations is already maintained in the solar
wind by the presence of weak spatial gradients (Spangler\cite{59}).

\begin{figure}[t]\vspace*{2mm}
  \centering{\includegraphics[width=5.42cm]{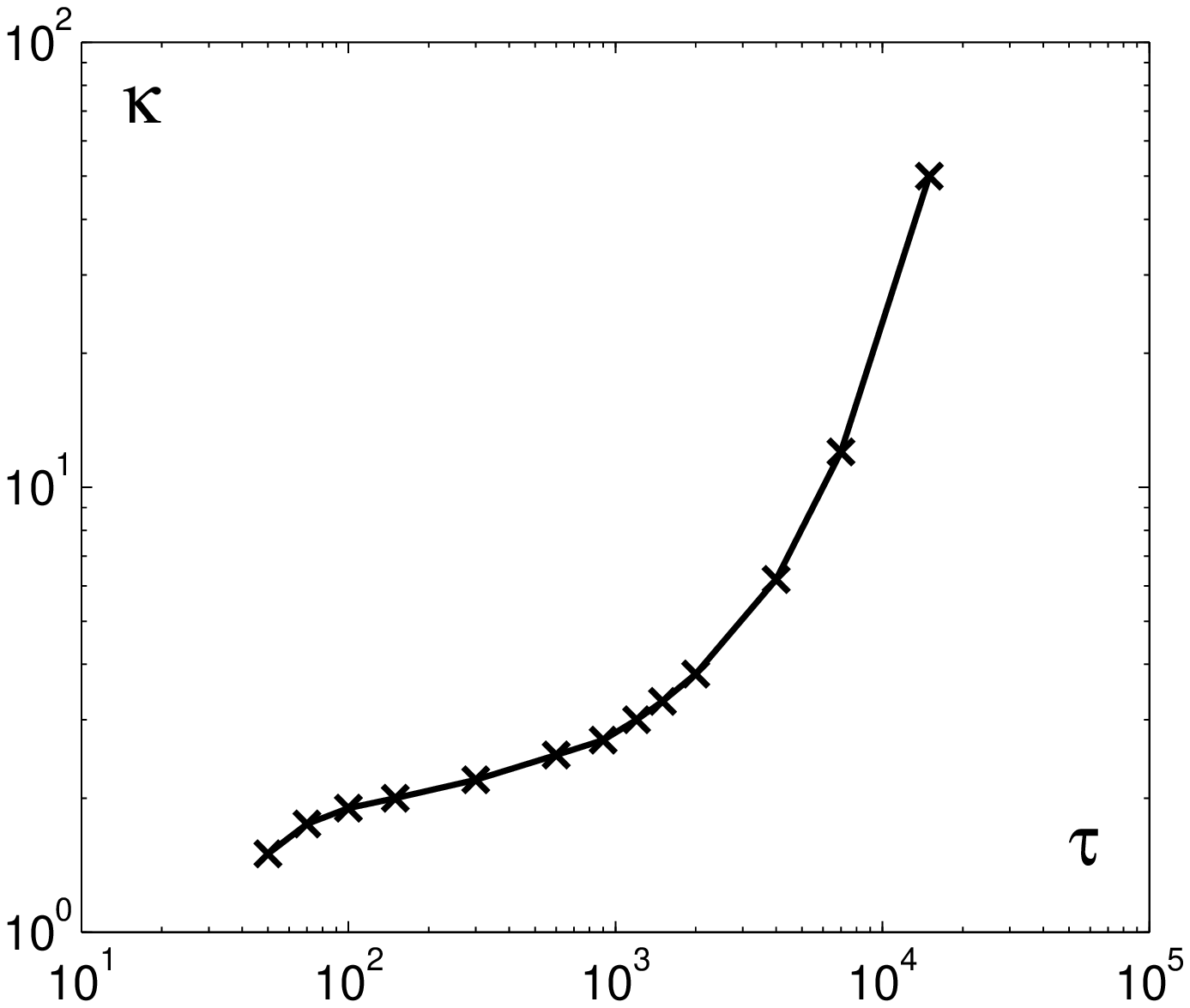}
            \includegraphics[width=5.5cm]{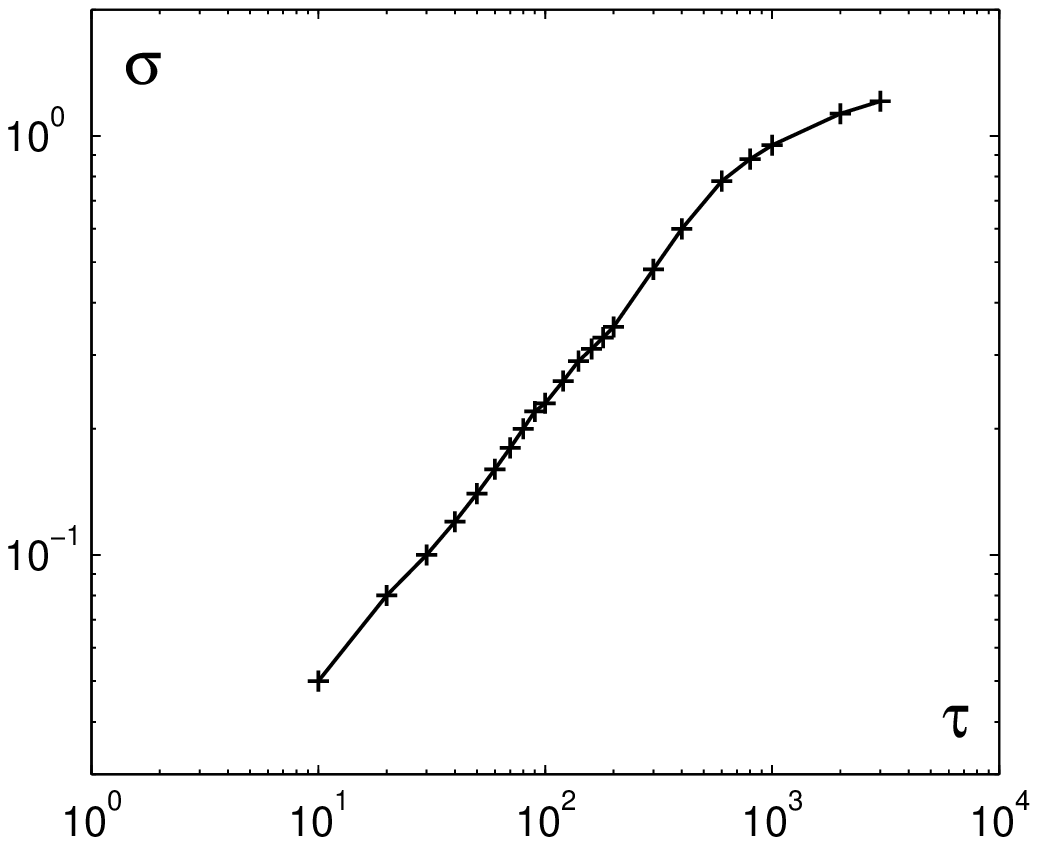}}
  \caption{\label{fig8}
The functional dependence of the spatial scale $\tau(\kappa)$ (left panel)
for slow speed and $\tau(\sigma)$ (right panel) for high speed conditions.}
\end{figure}

Fig. 8 (left panel) provides an estimation of the functional dependence
between the time lag $\tau$ and the nonextensive parameter
$\kappa$ for best fitting bi-kappa functions to the observed PDFs for slow
speed solar wind conditions ($v \leq 400 km s^{-1}$). As significant
global behavior the scale dependence of the PDFs for quiet conditions
appears to be independent of the variance or mean energy of the distribution.
The best fitting bi-kappa functions are found for a constant corresponding
parameter $\sigma$, indicating that the parameter $\kappa$, measuring
the degree of coupling within the system, governes primarily the
scale dependence of the PDF in the slow solar wind. As $\tau$
increases from small scales to the intermediate regime a pronounced plateau
formation is established, i.e. the relative increase in
$\kappa$-values with enhanced scales appears reduced. In other words, the
PDF shape appears at intermediate scales to be independent of the
spatial separation scale. Such a behavior may indicate the
presence of a transitional dynamical element characterizing a
balance between long and short-range interactions. Contrary, in high
speed streams ($v \ge  400 km s^{-1}$) best fits of bi-kappa functions
to the observed scale dependent PDFs are found when keeping $\kappa$
constant and varying only the parameter $\sigma$, left panel in Fig. 8.
Hence the scaling features in the fast wind appear independent of the
degree of coupling controlled by $\kappa$, but rely predominantly on
changes of the variance or mean energy.   
In summary, the theoretical nonextensive context indicates a significant and
physically contrary scaling behavior in slow and fast wind. For slow
solar wind conditions the correlations/intermittency
governed by $\kappa$ decrease with increasing scale, whereas the characteristic
energy governed by $\sigma$ remains constant. Contrary the scaling
properties in high speed streams are characterized by constant
correlations ($\kappa$) but enhanced variance with increasing scale.

$\kappa-$distributions reproduce the Maxwell-Boltzmann distribution for
$\kappa \rightarrow \infty$, a situation identifying $\kappa$ as an ordering
parameter that acounts for correlations within the system. 
Highly correlated turbulent conditions characterized by kappa distributions
represent stationary states far from equilibrium where a generalization of the
Boltzmann-Shannon entropy, as measure of the level of organization or
intermittency, applies (Goldstein and Lebowitz\cite{60}, Treumann\cite{61}).
Physically this can be understood considering a system at a certain nonlinear
stage where turbulence may reach a state of high energy level that is balanced
by turbulent dissipation. In this environment equilibrium statistics can be
extended to dissipative systems, approaching a stationary state beyond thermal
equilibrium (Gotoh and Kraichnan\cite{62}). Since turbulence
is driven in the solar wind by velocity shears we have choosen for the data analysis
intervals of low speed solar wind with limits in velocity space, where the
driving and dissipation conditions do not change significantly, maintaining
therefore the dynamical equilibrium condition. High speed intervals may depend
more on the occurence of dynamical processes on the Sun.

A multi-scale cascade mechanism is not the only way for a realization of
long range interactions. Let us provide a physical situation where large
and small scales are directly coupled and the nonlocal energy transfer is not
induced by cascading processes, following e.g. a log-normal model (Frisch\cite{63}).
An example is found in the context of MHD where Chang\cite{64} and 
Consolini and Chang\cite{65} proposed an intermittent turbulence model for the
solar wind and for the Earth's magnetotail, which  comprises neither cascades  nor
requires local interactions in Fourier space. In this scenario non-propagating
or convected fluctuations generate multiscale coherent structures (e.g. flux tubes),
which can interact,  deform and produce new sites of non-propagating fluctuations.
Coherent structures of the same polarity merge into a structure with lower local
energetic state, while structures of opposite polarities may repel each other.
This coherent structures can be considered as discrete interacting 'particles'
in MHD flows, responsible for the particular entropy within the system and
validating the analogy to the kinetic level of PDFs. Chang et al.\cite{66} have
computed PDFs of the intermittent fluctuations from direct numerical simulations
of interacting coherent structures. The resulting PDFs have typical leptokurtic
shapes, which can be well fitted again by a variety of models, including those
with predominantly local interactions in Fourier space. 
Since all models provide similar fitting accuaracy it is required to focus on
the underlying physical situation in turbulent flows. Hence, 
with regard to nonlocal interactions not based on cascade processes the
nonextensive entropy approach provides physically a justification for nonlocal
interactions and should therefore be favored over cascade models in such
processes. 

The entropy quantifies the degree of structuring in intermittent turbulence expressed
through singular multifractal measures, where also the parameter $\kappa$ (or $q$)
is related to the extremes of multifractal distributions (Lyra and Tsallis\cite{67},
Arimitsu and Arimitsu\cite{50}, Arimitsu and Arimitsu\cite{51}, Beck\cite{53}).
Since the nonextensive entropy approach is independent
of the mechanism leading to the structures  - in both situations, cascading processes
and  multiscale interacting coherent structures - or even in coexisting situations,
the entropy concept can be applied for the analysis and quantification of
resulting characteristics in turbulence. In summary, the majority of hitherto
existing models of intermittency in the solar wind essentially
correspond to the cascade picture of turbulence. Small-scale intermittency,
however, can be associated also by emerging topological complexity of coherent
sturctures in turbulence, which might be understood better through entropy
concepts, disregarding the goodness criteria of different fits. 

We provided specific examples based on multiscale interacting coherent
structures where the traditional cascade - and thus e.g. the log-normal approach
beside others -  should not be applied for physical reasons (not in terms of PDF
fitting accuracy). Therefore the proposed context based on entropy ganeralization
has a potential to describe the underlying physics suitably and thus justifies the
nonextensive approach. Certainly this must be viewed as an incomplete
concept in view of the complexity of intermittence in turbulence
(Cohen\cite{68}), in particular regarding the interactions with large scale structures.

Summarizing, a bi-kappa distribution family turns out theoretically
as consequence of the entropy generalization in nonextensive
thermo-statistics. The two-parameter global bi-kappa function
provides theoretically access to the scale dependence of the PDFs
observed in astrophysical plasma turbulence. The redistribution of a 
Gaussian on large scales into highly non-Gaussian leptokurtic and
long-tailed structures, manifest on small scales, is theoratically well
described by the family of nonextensive distributions.
Pseudo-additive entropy generalization provides the required physical
interpretation of the parameter $\kappa$ in terms of the degree of
nonextensivity of the system as a measure of nonlocality or coupling
due to long-range interactions whereas the variance $\sigma$ measures
the mean energy in the system. The scale dependence in the slow
speed solar wind is sensitive to variations of $\kappa$ and in high
speed streams to variations of $\sigma$. We argue that multiscale
coupling and intermittency of the turbulent solar wind fluctuations
must be related to the nonextensive character of the interplanetary
medium accounting for long-range interaction via the entropy generalization.

\end{document}